\documentclass[12pt,a4paper]{article} 
\usepackage{a4wide}

\title{Symmetric Calorons}
 \author{R.\ S.\ Ward\footnote{email: richard.ward@durham.ac.uk}
 \bigskip
\\Department of Mathematical Sciences,  \\ University of
Durham, \\Durham DH1 3LE}

\newcommand{\tr}{\mathop{\rm tr}\nolimits}
\newcommand{\half}{{\scriptstyle\frac{1}{2}}}
\newcommand{\RR}{{\bf R}}
\newcommand{\pa}{\partial}
\newcommand{\ii}{{\rm i}}
\newcommand{\ve}{\varepsilon}

\begin{document}

\maketitle
\abstract{\noindent Calorons (periodic instantons) interpolate between monopoles and
instantons, and their holonomy gives approximate Skyrmion configurations.  We show
that, for each caloron charge $N\leq4$, there exists a one-parameter family of
calorons which are symmetric under subgroups of the three-dimensional rotation
group.  In each family, the corresponding symmetric monopoles and symmetric
instantons occur as limiting cases.  Symmetric calorons therefore provide a
connection between symmetric monopoles, symmetric instantons and Skyrmions. 
}

\vskip 1truein
\noindent PACS 11.27.+d, 11.10.Lm, 11.15.-q

\newpage


\section{Introduction}

Calorons are finite-action self-dual gauge fields in four dimensions, which
are periodic in one of the four coordinates.  Call the periodic coordinate $t$,
with period $\beta$. Special cases include instantons on $\RR^4$ (where
$\beta\to\infty$) and BPS monopoles (where the gauge field is independent of $t$).
The holonomy $\Omega$ of the gauge field in the $t$-direction is a map from
$\RR^3$ to the gauge group, and as such can serve as an approximation to Skyrmions
\cite{AM89}.  Calorons therefore provide a link between monopoles, instantons and
Skyrmions.

Skyrmions resemble polyhedral shells, invariant under appropriate subgroups of the
three-dimensional rotation group O(3) \cite{BTC90,BS02}.  The idea of producing
approximate Skyrmion configurations as instanton holonomy has motivated several
studies of instantons invariant under such groups \cite{AM93,LM94,SS99,S03}.
Finally, there are symmetric monopoles \cite{HMM95} which have the same polyhedral
shape as the Skyrmions of corresponding charge, suggesting a kinship between Skyrmions
and monopoles \cite{HMS98}.  So symmetric calorons, namely calorons invariant under
subgroups $G$ of O(3) (rotations about the $t$-axis), are relevant in this context.
This Letter demonstrates the existence of symmetric calorons of charge $N$, for
$N\leq4$; they include, as limiting cases, symmetric monopoles and symmetric
instantons.

Large classes of calorons were described some years ago \cite{HS78,GPY81,A83,C87};
of these, only the $N=1$ case admits the relevant symmetry. So one needs more
general solutions.
There is a construction (the ADHMN construction) which generates caloron solutions
\cite{N84}, possibly all of them (see \cite{N03} for a recent analysis).  In the
last few years, this construction has been used to investigate and interpret
caloron solutions, especially those for which the holonomy $\Omega$ is non-trivial
at spatial infinity \cite{KB98,LL98,BB02}; but this recent work was not concerned with
symmetric solutions as such.  In this Letter, we shall see how symmetric calorons
arise from the ADHMN construction.


\section{Calorons, Monopoles and Skyrmions}

We take the gauge group to be SU(2) throughout.
The standard coordinates on $\RR^4$ are denoted
$x^\mu=(x^1,x^2,x^3,x^4)=(x^j,t)$; let $r$ be the quantity
defined by $r^2=x^j x^j$.  The gauge potential $A_{\mu}$ is anti-hermitian,
and the corresponding gauge field is
$F_{\mu\nu} = \pa_\mu A_\nu - \pa_\nu A_\mu +  [A_\mu,A_\nu]$.
A gauge transformation acts as
$A_\mu \mapsto \Lambda^{-1} A_\mu \Lambda + \Lambda^{-1} \pa_\mu \Lambda$.
A {\em caloron} \cite{HS78,GPY81,A83} is a gauge field
with the following properties:
\begin{itemize}
\item $A_{\mu}(x^\alpha)$ is periodic in $x^4=t$, with period $\beta$
      (in some gauge);
\item $A_{\mu}(x^\alpha)$ is smooth everywhere (in some gauge);
\item $F_{\mu\nu}$ is self-dual:
  $F_{\mu\nu}=\half\ve_{\mu\nu\alpha\beta}F_{\alpha\beta}$;
\item $\tr(F_{\mu\nu} F_{\mu\nu}) = O(1/r^4)$ as $r\to\infty$.
\end{itemize}
A special case of this is where $A_{\mu}$ is independent of $x^4=t$;
this is a {\em monopole}, where we make the usual interpretation of $A_t$
as a Higgs field $\Phi$.  The holonomy (or Wilson loop)
\begin{equation}\label{hol}
 \Omega(x^j)={\cal P}\exp\left[-\int_0^{\beta}A_t(x^j,t)\,dt\right]
\end{equation}
in the $t$-direction takes values in the gauge group; under a periodic
gauge transformation, it transforms as
\begin{equation}\label{hol_gauge}
  \Omega(x^j)\mapsto
      \Lambda(x^j,0)^{-1}\, \Omega(x^j)\, \Lambda(x^j,0).
\end{equation}
The quantity $\Omega(x^j)$ is, in general, non-trivial at spatial
infinity \cite{GPY81}; but for the examples below,
$\Omega(x^j)$ tends to a constant group element
(in fact the identity) as $r\to\infty$.  Such a field may be viewed as an
approximate {\em Skyrmion}
configuration; the Skyrmion number is the degree of $\Omega$, and the
normalized Skyrme energy is
\begin{equation}\label{SkyrmeEn}
  E = \frac{1}{12\pi^2}\int\left\{ -\half\tr(L_jL_j)
   -{{\scriptstyle\frac{1}{16}}}\tr([L_i,L_j][L_i,L_j])\right\}\,d^3x,
\end{equation}
where $L_j=\Omega^{-1}\pa_j\Omega$.  Provided $\Omega$ is asymptotically
trivial, the topological charge (caloron number)
\begin{equation}\label{top_charge}
   N = -\frac{1}{32\pi^2}\int_0^{\beta}dt\int\,d^3x
       \,\tr\left(\ve_{\mu\nu\alpha\beta}F_{\mu\nu}F_{\alpha\beta}\right)
\end{equation}
is an integer, and is equal to the Skyrmion number of $\Omega$
\cite{GPY81}.  In the $t$-independent (monopole) case, it is also the monopole
number, provided we take $\beta$ to be related to the asymptotic norm
of the Higgs field by
\begin{equation}\label{asymp_Higgs}
-\frac{1}{2}\tr\left(\Phi_{\infty}\right)^2=\left(\frac{\pi}{\beta}\right)^2.
\end{equation}

A large number of caloron solutions can be generated \cite{HS78} by the
Corrigan-Fairlie-'tHooft \cite{CF77} or Jackiw-Nohl-Rebbi \cite{JNR77} ansatz.
These express the gauge potential in terms of a solution $\phi$
(periodic, in the caloron case) of
the four-dimensional Laplace equation.  For example, the component $A_t$ is
given by
\begin{equation}\label{ansatz}
  A_t = \frac{\ii}{2} \left(\pa_j\log\phi\right)\sigma_j
\end{equation}
where $\sigma_j$ are the Pauli matrices.  For the JNR solutions one has
$\phi\to0$ as $r\to\infty$, whereas for the CF'tH solutions one has
$\phi\to1$ as $r\to\infty$.  In the case of instantons on $\RR^4$, one
regards the CF'tH solutions as being limiting cases of the JNR
solutions,  but for calorons it is the other way round: to produce an
$N$-caloron in JNR form, one uses a $\phi$ with $N$ poles (not $N+1$ as
for instantons), and this is a limiting case of the CF'tH form with $N$
poles.

To illustrate this, let us review the $N=1$ case.  The 1-caloron (with
trivial holonomy at infinity) is generated \cite{HS78} by the 1-pole function
\begin{equation}\label{phi}
  \phi = 1 + \frac{W^2\sinh(\mu r)}{2r[\cosh(\mu r)-\cos(\mu t)]},
\end{equation}
where $\mu=2\pi/\beta$, and $W>0$ is a constant.  This caloron is
spherically-symmetric; it depends on the period $\beta$ and on the
parameter $W$.  The gauge field is not affected by an overall scale factor
in $\phi$, so the $W\to\infty$ limit of (\ref{phi}) gives, in effect, the
JNR-type solution with
\begin{equation}\label{phi_JNR}
  \phi = \frac{\sinh(\mu r)}{2r[\cosh(\mu r)-\cos(\mu t)]};
\end{equation}
this corresponds to a 1-caloron which is in fact gauge-equivalent to the
1-monopole \cite{R79}.  Another way of viewing things is to use the
dimensionless combination $\theta=\beta/W^2$: for $\theta=0$ (or $W\to\infty$)
we get the 1-monopole, while for $\theta\to\infty$ (or $\beta\to\infty$) we get the
1-instanton on $\RR^4$.  In other words, we have a one-parameter family of 
spherically-symmetric calorons,
with the 1-monopole at one end and the 1-instanton at the other end.
The holonomy $\Omega(x^j)$ can be computed exactly in this case \cite{EK89,NZ89};
if one restricts to spherically-symmetric gauges, then $\Omega$ is actually
gauge-invariant. 
The Skyrme energy (\ref{SkyrmeEn}) of this configuration $\Omega$ attains a minimum
for $\theta\approx7$; this minimum is only slightly less \cite{EK89} than the
value obtained from 1-instanton holonomy.

It is straightforward to produce spherically-symmetric calorons of higher charge
in this way: for example, the function
\begin{equation}\label{phi2}
  \phi = 1 + \frac{W^2\sinh(\mu r)}{2r[\cosh(\mu r)-\cos(\mu t)]}
           + \frac{{\widehat W}^2\sinh(\mu r)}{2r[\cosh(\mu r)-\cos(\mu(t-t_0))]}
\end{equation}
generates a spherically-symmetric 2-caloron, for any $t_0\in(0,\beta)$ and
$W,{\widehat W}>0$.
The holonomy of this is a spherically-symmetric (hedgehog) 2-Skyrmion
configuration (cf.\ \cite{AM89,AM93}).  The limits $\beta\to\infty$ and
$W,{\widehat W}\to\infty$ are both regular; the former is a 2-instanton,
but the latter is not a 2-monopole (since, unlike in the $N=1$ case, the
$t$-dependence cannot be gauged away).
It seems very unlikely that the CF'tH ansatz can yield any examples (other than
for $N=1$) of symmetric calorons having symmetric monopoles as a limiting
case --- for that, one needs more general solutions.  A way of generating
such solutions is described in the next section.

\section{The ADHMN Construction for Calorons}

There is a construction which produces caloron solutions \cite{N84};
for gauge group SU(2), and for calorons which have trivial holonomy at
infinity, it is as follows.  As before, $N$ is a postive integer which
will turn out to be the caloron charge, and $\beta$ is a positive number
which will turn out to be the caloron period.  It is convenient to use
quaternion notation, with a quaternion $q$ being represented by the $2\times2$
matrix $q^4 + \ii q^j\sigma^j$; in particular, $x^\mu$ corresponds to
the quaternion $x=t + \ii x^j\sigma^j$.  The unit quaternion ($q^4=1$, $q^j=0$)
is denoted {\bf1}.

The Nahm data consists of four hermitian $N\times N$ matrix functions
$T_{\mu}(s)$, and an $N$-row-vector $W$ of quaternions, such that $T_{\mu}(s)$
is periodic in the real variable $s$ with period $2\pi/\beta$, and the Nahm
equation
\begin{equation}\label{Nahm_cal}
 \frac{d}{ds} T_j - \ii[T_4,T_j]-\frac{\ii}{2}\ve_{jkl}[T_k,T_l]
   = \frac{1}{2} \tr_2\left(\sigma_j W^{\dagger}\,W\right)\delta(s-\pi/\beta)
\end{equation}
is satisfied.  The trace is over quaternions, so the right-hand-side is
an $N\times N$ hermitian matrix (as is the left-hand-side).  
Given such data, we construct a caloron as follows.  Let $U(s,x)$ be an
$N$-column-vector of quaternions, and $V(x)$ a single quaternion, such that
\begin{enumerate}
\item $U(s,x)$ is periodic in $s$ with period $2\pi/\beta$;
\item $U(s,x+\beta)=U(s,x)\exp(\ii\beta s)$;
\item $V(x+\beta)=V(x)$;
\item $\int_{-\pi/\beta}^{\pi/\beta}U(s,x)^{\dagger}\,U(s,x)\,ds
        + V(x)^{\dagger}\,V(x)={\bf1}$;
\item $U$ and $V$ satisfy the linear equation
\begin{equation}\label{Nahm_cal_lin}
 \frac{d}{ds}U - \left[\ii(T_4+tI_n)\otimes{\bf1}+I_n\otimes x^j\sigma^j
  + T_j\otimes\sigma^j\right]U = \ii\, W^{\dagger}\, V\delta(s-\pi/\beta).
\end{equation}
\end{enumerate}
Note that both $T_j$ and $U$ are periodic in $s$, and have jump discontinuities
at one value of $s$, which we have taken to be $s=\pi/\beta$.  The
discontinuities could equally well be located anywhere else; the choice in
(\ref{Nahm_cal}) and (\ref{Nahm_cal_lin}) is for later convenience.  Note also
that the overall quaternionic phase of the $N$-vector $W=[W_1 \ldots W_N]$
is irrelevant; so we may, without loss of generality, take $W_1$ to be real.

The pair $(U,V)$ determines the caloron gauge potential according to
\begin{equation}\label{Nahm_cal_A}
  A_{\mu}= V(x)^{\dagger}\,\pa_{\mu}V(x) +
 \int_{-\pi/\beta}^{\pi/\beta}U(s,x)^{\dagger}\,\pa_{\mu}U(s,x)\,ds.
\end{equation}
The freedom in $(U,V)$ is $U\mapsto U\Lambda$, $V\mapsto V\Lambda$, where
$\Lambda$ is a quaternion satisfying $\Lambda^{\dagger}\,\Lambda=1$; this
corresponds exactly to the gauge freedom in $A_{\mu}$.

By contrast, the usual formulation of the ADHMN construction for monopoles
involves three matrices $T_j(s)$, satisfying
\begin{equation}\label{Nahm_mon}
 \frac{d}{ds} T_j -\frac{\ii}{2}\ve_{jkl}[T_k,T_l] = 0.
\end{equation}
In this case, the $T_j(s)$ are not periodic in $s$, but rather are smooth on
the open interval $|s|<1$, with poles at the endpoints $s=\pm1$.  (The length
of this interval sets the scale of the monopole.)  In addition, the $T_j$
satisfy
\begin{equation}\label{Nahm_mon2}
  T_j(-s) = T_j(s)^t.
\end{equation}
The idea here is that given a solution of the monopole Nahm equation
(\ref{Nahm_mon}), one may re-interpret it as a solution of the caloron
Nahm equation (\ref{Nahm_cal}), with $T_4=0$ and with a suitable choice of $W$,
namely such that
\begin{equation}\label{W1}
  T_j(-\pi/\beta) - T_j(\pi/\beta) =
    \frac{1}{2} \tr_2\left(\sigma_j W^{\dagger}\,W\right).
\end{equation}
We need to take $\beta>\pi$, so that the $T_j$ are bounded for
$|s|\leq\pi/\beta$.  The symmetric part of $T_j$ can, because of
(\ref{Nahm_mon2}), be regarded as a continuous periodic function on
$[-\pi/\beta,\pi/\beta]$; while the antisymmetric part of $T_j$ has
a jump discontinuity as in (\ref{W1}).

The limit $\beta\to\pi$ is the original monopole, while
the limit $\beta\to\infty$ gives an instanton on $\RR^4$.  This instanton
limit works as follows.
For $\beta\gg\pi$, we are solving (\ref{Nahm_cal_lin}) on the small interval
$|s|\leq\pi/\beta$, so we may approximate the solution as $U(s)=U_0+U_1s$.
Equation (15) then gives
\begin{equation}\label{limit1}
  U_1 = (it+x^j\sigma^j+{\cal T}_j\otimes\sigma^j)U_0
       = -\frac{\ii\beta}{2\pi}W^{\dagger}\,V,
\end{equation}
where ${\cal T}_j=T_j(0)$, and where $U_0$ and $V$ satisfy the constraint
\begin{equation}\label{limit2}
U_0^{\dagger}\,U_0 + V(x)^{\dagger}\,V(x)={\bf1}.
\end{equation}
If we write $\Lambda=\sqrt{\beta/2\pi}\,W$, then this is exactly the ADHM
construction \cite{ADHM78} for instantons, with the ADHM matrix $\Delta$
being given by
\begin{equation}\label{limit3}
  \Delta = \left[\begin{array}{c}
                 \Lambda\\
                x + \ii{\cal T}_j\otimes\sigma^j
           \end{array}\right].
\end{equation}
This $\Delta$ is an $(n+1)\times n$ matrix of quaternions, satisfying the
condition that $\Delta^{\dagger}\,\Delta$ is an $n\times n$ real matrix.

Let us now consider calorons which are symmetric under subgroups of the
three-dimensional rotation group acting on $x^j$.  For any rotation $R$,
let $R_2\in\,\,$SU(2) denote the image of $R$ in the 2-dimensional irreducible
representation of SO(3); in other words, $R$ acts on the quaternion $x$
according to $x\mapsto R_2^{-1}xR_2$.  Similarly, let $R_N$ denote the image
of $R$ in the $N$-dimensional irreducible representation of SO(3), and write
$\Theta_R=R_N\otimes R_2$.  A monopole is invariant \cite{HMM95} under the
group $G\subseteq\,\,$SO(3) iff
\begin{equation}\label{mon_sym}
 \Theta_R^{-1} (T_j\otimes\sigma^j)\, \Theta_R = T_j\otimes\sigma^j
\end{equation}
for all $R\in G$.  For the corresponding caloron to be $G$-invariant, we need
an additional condition on $W$, and this is easily seen (from (\ref{Nahm_cal})
and (\ref{Nahm_cal_lin})) to be
\begin{equation}\label{W2}
 \Theta_R W^{\dagger} = W^{\dagger}\,\tau_R
\end{equation}
where $\tau_R$, for each $R\in G$, is some quaternionic phase (namely
a quaternion
with $\tau_R^{\dagger}\,\tau_R={\bf1}$).  So given a symmetric monopole,
there is a family of symmetric calorons parametrized by the solutions $W$
(if there are any) of (\ref{W1}) and (\ref{W2}).  In the $N=1$ case, for
example, we have $G=\,$SO(3) (spherical symmetry) and $T_j=0$; and $W$ is
an arbitrary positive constant, which is precisely the parameter appearing
in the expression (\ref{phi}).  In the next section, we shall see that
analogous one-parameter families of symmetric calorons exist for $N=2$,
$3$ and $4$.


\section{Symmetric Examples for $N=2,3,4$}

We begin with the $N=2$ case, taking $G=\,$SO(2) (corresponding to
rotations about the $x^2$-axis).  The solution of (\ref{Nahm_mon})
which generates the axially-symmetric $N=2$ monopole is
$T_j(s)=f_j(s)\sigma_j$ (not summed over $j$), where
\begin{equation}\label{Nahm_2mon}
 f_1=f_3=\frac{\pi}{4}\sec(\pi s/2), \quad  f_2=-\frac{\pi}{4}\tan(\pi s/2).
\end{equation}
Then (\ref{W1}) and (\ref{W2}) have a solution $W$ which is unique (given
that $W_1$ is real), namely
\begin{equation}\label{W_cal2}
  W=\lambda[{\bf1} \quad -\ii\sigma_2], \quad{\rm where}\quad
    \lambda=\sqrt{\frac{\pi}{2}\tan\left(\frac{\pi^2}{2\beta}\right)}.
\end{equation}
So we get a family of $N=2$ axially-symmetric caloron solutions, depending
on the parameter $\beta>\pi$. It is possible to solve (\ref{Nahm_cal_lin})
analytically, and hence obtain exact expressions for the caloron (cf.\ \cite{P83}
for the monopole case), although the expressions are rather complicated. 
The limit $\beta\to\pi$ is the 2-monopole, and $\beta\to\infty$ is a
2-instanton on $\RR^4$, generated by the ADHM matrix
\begin{equation}\label{inst2}
  \Delta = \frac{\pi}{4}\left[\begin{array}{cc}
                  \sqrt{2}    & -\ii\sqrt{2}\sigma_2\\
                  \ii\sigma_3 & \ii\sigma_1\\
                  \ii\sigma_1 & -\ii\sigma_3
           \end{array}\right]
          +\left[\begin{array}{cc}
                  0 & 0\\
                  x & 0\\
                  0 & x
           \end{array}\right].
\end{equation}
This axially-symmetric 2-instanton can be obtained in the JNR
form, and its holonomy was used to approximate the
minimum-energy 2-Skyrmion \cite{HGOA90,AM93}.
The holonomy $\Omega$ of the caloron gives a one-parameter family of
axially-symmetric 2-Skyrmion configurations; as in the $N=1$ case, this
gives an approximation to the true Skyrmion which is better than the instanton
one, but only marginally so.

Let us now consider the $N=3$ case.  There is a 3-monopole with tetrahedral
symmetry \cite{HMM95,HS96}, corresponding to the following Nahm data. (Note that
the $T_j$ in \cite{HMM95,HS96} have to be multiplied by a factor of $-\ii$ to
agree with the conventions used here.)  Define
\begin{equation}\label{tet1}
  \Sigma_1 = 2\ii\left[\begin{array}{ccc}
                  0 & 0 & 0\\
                  0 & 0 & -1\\
                  0 & 1 & 0
           \end{array}\right],\quad
  \Sigma_2 = 2\ii\left[\begin{array}{ccc}
                  0 & 0 & 1\\
                  0 & 0 & 0\\
                  -1 & 0 & 0
           \end{array}\right],\quad
  \Sigma_3 = 2\ii\left[\begin{array}{ccc}
                  0 & -1 & 0\\
                  1 & 0 & 0\\
                  0 & 0 & 0
           \end{array}\right],
\end{equation}
and
\begin{equation}\label{tet2}
  S_1 = \left[\begin{array}{ccc}
                 0 & 0 & 0\\
                 0 & 0 & 1\\
                 0 & 1 & 0
           \end{array}\right],\quad
  S_2 = \left[\begin{array}{ccc}
                  0 & 0 & 1\\
                  0 & 0 & 0\\
                  1 & 0 & 0
           \end{array}\right],\quad
  S_3 = \left[\begin{array}{ccc}
                  0 & 1 & 0\\
                  1 & 0 & 0\\
                  0 & 0 & 0
           \end{array}\right].
\end{equation}
Then $T_j(s)=x(s)\Sigma_j+y(s)S_j$, where
\begin{equation}\label{tet3}
  x(s)=-\frac{\omega\, \wp'(u)}{12\,\wp(u)}, \qquad
     y(s)=-\frac{\omega}{\sqrt{3}\,\wp(u)},
\end{equation}
with $u=\omega(s+3)/3$ and $\omega=\Gamma(1/6)\Gamma(1/3)/(4\sqrt{\pi})$.
Here $\wp$ is the Weierstrass p-function satisfying $\wp'(u)^2=4\wp(u)^3+4$.
The unique solution of (\ref{W1}), with $W_1>0$, is
\begin{equation}\label{W_cal3}
  W=\lambda[{\bf1} \quad \ii\sigma_3 \quad -\ii\sigma_2],\ {\rm where}\
    \lambda=2\sqrt{x(\pi/\beta)}.
\end{equation}
Explicit calculation then verifies that (\ref{W2}) is satisfied for each
of the elements of the tetrahedral group.  So we have a one-parameter family
of tetrahedrally-symmetric 3-calorons, interpolating between the tetrahedral
3-monopole and a tetrahedrally-symmetric 3-instanton.  The latter
is generated by the ADHM matrix
\begin{equation}\label{inst3}
  \Delta = \frac{\omega}{\sqrt{3}}\left[\begin{array}{ccc}
                  {\bf1} & \ii\sigma_3 & -\ii\sigma_2\\
                       0 & \ii\sigma_3  & \ii\sigma_2\\
                       \ii\sigma_3 & 0 & \ii\sigma_1\\
                       \ii\sigma_2 & \ii\sigma_1 & 0
           \end{array}\right]
          +\left[\begin{array}{ccc}
                0 & 0 & 0\\
                x & 0 & 0\\
                0 & x & 0\\
                0 & 0 & x
           \end{array}\right].
\end{equation}
A tetrahedrally-symmetric 3-instanton can also be obtained in JNR
form, and its holonomy was used to approximate the minimum-energy
3-Skyrmion \cite{LM94}.

For the final example, we consider 4-calorons with cubic symmetry (so $G$ is
the 24-element octahedral group).  The Nahm data
in \cite{HMM95} and \cite{HS96} do not satisfy (\ref{Nahm_mon2}), and so we have
to change to a basis in which (\ref{Nahm_mon2}) holds.  Define
\[
  \Sigma_1 = -\left[\begin{array}{cccc}
                 -\sqrt{3} & 0 & -\ii & -1\\
                  0 & \sqrt{3} & -1 & \ii\\
                  \ii & -1 & -\sqrt{3} & 0\\
                  -1 & -\ii & 0 & \sqrt{3}
           \end{array}\right],\quad
  \Sigma_2 = -\left[\begin{array}{cccc}
                  0 & \sqrt{3} & 1 & -\ii\\
                  \sqrt{3} & 0 & -\ii & -1\\
                  1 & \ii & 0 & \sqrt{3}\\
                  \ii & -1 & \sqrt{3} & 0
           \end{array}\right],
\]
\[
  \Sigma_3 = -\left[\begin{array}{cccc}
                  2 & -\ii & 0 & 0\\
                  \ii & 2 & 0 & 0\\
                  0 & 0 & -2 & -\ii\\
                  0 & 0 & \ii & -2
           \end{array}\right],\quad
  S_1 = -2\left[\begin{array}{cccc}
                  \sqrt{3} & 0 & -4\ii & 1\\
                  0 & -\sqrt{3} & 1 & 4\ii\\
                  4\ii & 1 & \sqrt{3} & 0\\
                  1 & -4\ii & 0 & -\sqrt{3}
           \end{array}\right],
\]
\[
  S_2 = -2\left[\begin{array}{cccc}
                  0 & -\sqrt{3} & -1 & -4\ii\\
                  -\sqrt{3} & 0 & -4\ii & 1\\
                  -1 & 4\ii & 0 & -\sqrt{3}\\
                   4\ii & 1 & -\sqrt{3} & 0
           \end{array}\right],\quad
  S_3 = -4\left[\begin{array}{cccc}
                  -1 & -2\ii & 0 & 0\\
                  2\ii & -1 & 0 & 0\\
                  0 & 0 & 1 & -2\ii\\
                  0 & 0 & 2\ii & 1
           \end{array}\right].
\]
Then $T_j(s)=x(s)\Sigma_j+y(s)S_j$, where
\begin{equation}\label{cube3}
  y = \frac{\omega_2}{10\,\wp'(u)}, \qquad
  x = [5\wp(u)^2-3]\,y,
\end{equation}
with $\omega_2=(1+\ii)\,\Gamma(1/4)^2/(4\sqrt{2\pi})$ and $u=\omega_2(s+1)/2$.
Here $\wp$ is the Weierstrass p-function satisfying $\wp'(u)^2=4\wp(u)^3-4\wp(u)$.
The condition (\ref{Nahm_mon2}) follows from the relations
\begin{equation}\label{cube4}
  \left[\begin{array}{c} x(-s)\\ y(-s)\end{array}\right]
   = \frac{1}{5}\left[\begin{array}{cc}
                3 & -16\\ -1 & -3\end{array}\right]
       \left[\begin{array}{c} x(s)\\ y(s)\end{array}\right].
\end{equation}
Then, as before, (\ref{W1}) has a unique solution
\begin{equation}\label{W_cal4}
  W=\lambda[{\bf1} \quad \ii\sigma_3 \quad \ii\sigma_1 \quad \ii\sigma_2],\
   {\rm where}\ \lambda=\sqrt{2\,x(\pi/\beta)+16\,y(\pi/\beta)};
\end{equation}
and one may check explicitly that (\ref{W2}) is satisfied for each
element of the octahedral group.  So here we have a one-parameter family
of octahedrally-symmetric 4-calorons, interpolating between the cubic
(octahedrally-symmetric) 4-monopole and an octahedrally-symmetric 4-instanton.
This instanton is generated by the ADHM matrix
\begin{equation}\label{inst4}
  \Delta = \frac{|\omega_2|}{\sqrt{2}}\left[\begin{array}{cccc}
                  {\bf1} & \ii\sigma_3 & \ii\sigma_1 & \ii\sigma_2\\
   \frac{\sqrt{3}}{2}\ii\sigma_1-\ii\sigma_3 & -\frac{\sqrt{3}}{2}\ii\sigma_2 & -\frac{1}{2}\ii\sigma_2 & \frac{1}{2}\ii\sigma_1\\
   -\frac{\sqrt{3}}{2}\ii\sigma_2 & -\frac{\sqrt{3}}{2}\ii\sigma_1-\ii\sigma_3 & \frac{1}{2}\ii\sigma_1 & \frac{1}{2}\ii\sigma_2\\
   -\frac{1}{2}\ii\sigma_2 & \frac{1}{2}\ii\sigma_1 & \frac{\sqrt{3}}{2}\ii\sigma_1+\ii\sigma_3 & -\frac{\sqrt{3}}{2}\ii\sigma_2\\
   \frac{1}{2}\ii\sigma_1 & \frac{1}{2}\ii\sigma_2 & -\frac{\sqrt{3}}{2}\ii\sigma_2 & -\frac{\sqrt{3}}{2}\ii\sigma_1+\ii\sigma_3
           \end{array}\right]
          +\left[\begin{array}{cccc}
                0 & 0 & 0 & 0\\
                x & 0 & 0 & 0\\
                0 & x & 0 & 0\\
                0 & 0 & x & 0\\
                0 & 0 & 0 & x
           \end{array}\right],
\end{equation}
which may be compared with the symmetric 4-instanton example described
in \cite{LM94}.

In conclusion,
we have seen that, at least for charge $N\leq4$, there is an intimate
connection between symmetric monopoles, symmetric calorons, symmetric
instantons, and (via holonomy) Skyrmions.  Many open questions remain,
of which the following are a few.
\begin{itemize}
 \item Several more symmetric monopoles (of higher charge) are known --- do
  all of these arise as limiting cases of calorons with the same symmetry?
  More generally, is it true that any symmetric monopole has to be a
  special case of a symmetric caloron?
 \item Similarly, does every symmetric instanton \cite{SS99} extend to
  a family of symmetric calorons?  Note that such families are much
  more general, in that there may not be a symmetric monopole at the
  `other end'.
 \item What is the role of harmonic maps, which are known to be related
  to symmetric monopoles and Skyrmions \cite{HMS98}?  Does this involve
  the interpretation of calorons as monopoles with a loop group as their
  gauge group \cite{GM88,N00}?
\end{itemize}



\newpage

\begin{thebibliography}{99}

\bibitem{AM89}
M~F~Atiyah and N~S~Manton, Skyrmions from instantons.
         {\it Phys Lett B} {\bf222} (1989) 438--442.

\bibitem{BTC90}
E~Braaten, S~Townsend and L~Carson,  Novel structure of static multisoliton
  solutions in the Skyrme model. {\it Phys Lett B} {\bf235} (1990) 147--152.

\bibitem{BS02}
R~Battye and P~M~Sutcliffe, Skyrmions, fullerenes and rational maps.
           {\it Rev Math Phys} {\bf14} (2002) 29--85.

\bibitem{AM93}
M~F~Atiyah and N~S~Manton, Geometry and kinematics of two Skyrmions.
       {\it Commun Math Phys} {\bf152} (1993) 391--422.

\bibitem{LM94}
R~A~Leese and N~S~Manton, Stable instanton-generated Skyrme fields with
    baryon numbers three and four. {\it Nucl Phys A} {\bf572} (1995) 575--599.

\bibitem{SS99}
M~A~Singer and P~M~Sutcliffe, Symmetric instantons and Skyrme fields.
   {\it Nonlinearity} {\bf12} (1999) 987--1003.

\bibitem{S03}
P~M~Sutcliffe, Instantons and the buckyball.  hep-th/0309157

\bibitem{HMM95}
N~J~Hitchin, N~S~Manton and M~K~Murray, Symmetric monopoles.
       {\it Nonlinearity} {\bf8} (1995) 661--692.

\bibitem{HMS98}
C~J~Houghton, N~S~Manton and P~M~Sutcliffe, Rational maps, monopoles and Skyrmions.
      {\it Nucl Phys B} {\bf510} (1998) 507--537.

\bibitem{HS78}
B~J~Harrington and H~K~Shepard, Periodic Euclidean solutions and the
    finite-temperature Yang-Mills gas.
    {\it Phys Rev D} {\bf17} (1978) 2122--2125.

\bibitem{GPY81}
D~J~Gross, R~D~Pisarski and L~G~Yaffe, QCD and instantons at finite
    temperature.  {\it Rev Mod Phys} {\bf53} (1981) 43--80.

\bibitem{A83}
A~Actor, Self dual solutions of the temperature SU(2) Yang-Mills theory.
         {\it Ann Phys} {\bf148} (1983) 32--56.

\bibitem{C87}
A~Chakrabarti, Periodic generalizations of static, self-dual SU(2) gauge fields.
      {\it Phys Rev D} {\bf35} (1987) 696--706.

\bibitem{N84}
W~Nahm, Self-dual monopoles and calorons. {\it Springer Lecture Notes in
   Physics} {\bf201} (1984) 189--200.

\bibitem{N03}
T~M~W~Nye, The geometry of calorons. hep-th/0311215

\bibitem{KB98}
T~C~Kraan and P~van~Baal, Periodic instantons with non-trivial holonomy.
          {\it Nucl Phys B} {\bf533} (1998) 627--659.

\bibitem{LL98}
K~Lee and C~Lu, SU(2) calorons and magnetic monopoles.
         {\it Phys Rev D} {\bf58} (1998) 025011.

\bibitem{BB02}
F~Bruckmann and P~van~Baal, Multi-caloron solutions.
          {\it Nucl Phys B} {\bf645} (2002) 105--133.

\bibitem{CF77}
E~Corrigan and D~B~Fairlie, Scalar field theory and exact solutions to a classical
     SU(2) gauge theory. {\it Phys Lett B} {\bf67} (1977) 69--71.

\bibitem{JNR77}
R~Jackiw, C~Nohl and C~Rebbi, Conformal properties of pseudoparticle
    configurations.  {\it Phys Rev D} {\bf15} (1977) 1642--1646.

\bibitem{R79}
P~Rossi, Propagation functions in the field of a monopole.
         {\it Nucl Phys B} {\bf149} (1979) 170--188.

\bibitem{EK89}
K~J~Eskola and  K~Kajantie, Thermal Skyrmion-like configuration.
   {\it Z Phys C} {\bf44} (1989) 347--348.

\bibitem{NZ89}
M~A~Novak and I~Zahed, Skyrmions from instantons at finite temperature.
         {\it Phys Lett B} {\bf230} (1989) 108--112.

\bibitem{ADHM78}
M~F~Atiyah, V~G~Drinfeld, N~J~Hitchin and Yu~I~Manin, Construction of
    instantons. {\it Phys Lett A} {\bf65} (1978) 185--187.

\bibitem{P83}
H~Panagopoulos, Multimonopoles in arbitray gauge groups and the complete
    SU(2) two-monopole system.  {\it Phys Rev D} {\bf28} (1989) 380--384.

\bibitem{HGOA90}
A~Hosaka, S~M~Griffies, M~Oka and R~D~Amado, Two skyrmion interaction for the
    Atiyah-Manton ansatz.  {\it Phys Lett B} {\bf251} (1990) 1--5.

\bibitem{HS96}
C~J~Houghton and P~M~Sutcliffe, Tetrahedral and cubic monopoles.
      {\it Commun Math Phys} {\bf180} (1996) 343--361.

\bibitem{GM88}
H~Garland and M~K~Murray, Kac-Moody monopoles and periodic instantons.
        {\it Commun Math Phys} {\bf120} (1988) 335--351.

\bibitem{N00}
P~Norbury, Periodic instantons and the loop group.
      {\it Commun Math Phys} {\bf212} (2000) 557--569.

\end{thebibliography}
\end{document}